\documentclass[twocolumn,showpacs,preprintnumbers,amsmath,amssymb]{revtex4}

\usepackage{graphicx}
\usepackage{dcolumn}
\usepackage{bm}

\begin{document}

\preprint{ }

\title{Circular polarization memory effect
in enhanced backscattering of light \\ under partially coherent
illumination}

\author{Young L. Kim}
\altaffiliation {Young Kim's email address is younglae@northwestern.edu}
\affiliation {Department of Preventive Medicine, Northwestern University, Chicago, IL 60611}%
\author{Prabhakar Pradhan}
\author{Min H. Kim}
\author{Vadim Backman}

\affiliation{Biomedical Engineering Department, Northwestern
University,
Evanston, IL 60208}%


\date{\today}

\begin{abstract}
We experimentally study the propagation of circularly polarized
light in the sub-diffusion regime by exploiting enhanced
backscattering (EBS, also known as coherent backscattering) of
light under low spatial coherence illumination.  We demonstrate
for the first time that circular polarization memory effect exists
in EBS over a large range of scatterers' sizes in this regime.  We
show that EBS measurements under low spatial coherence
illumination from the helicity preserving and orthogonal helicity
channels cross over as the mean free pathlength of light in media
varies, and that the cross point indicates the transition from
multiple to double scattering in EBS of light.

\end{abstract}

\pacs{42.25.Dd, 42.25.Kb, 42.25.Ja.}
\maketitle

The circular polarization memory effect is an unexpected
preservation of the initial helicity (or handedness) of circular
polarization of multiply scattered light in scattering media
consisting of large particles.  Mackintosh \textit{et al}. [1]
first observed that the randomization of the helicity required
unexpectedly far more scattering events than did the randomization
of its propagation in media of large scatterers.  Bicout
\textit{et al}. [2] demonstrated that the memory effect can be
shown by measuring the degree of circular polarization of
transmitted light in slabs. Using numerical simulations of vector
radiative transport equations, Kim and Moscoso~[3] explained the
effect as the result of successive near-forward scattering events
in large scatterers. Recently, Xu and Alfano~[4] derived a
characteristic length of the helicity loss in the diffuse regime
and showed that this characteristic length was greater than the
transport mean free pathlength $l_s^*$ for the scatterers of large
sizes.  Indeed, the propagation of circularly polarized light in
random media has been investigated mainly using either numerical
simulations or experiments in the diffusion regime, in part
because its experimental investigation in the sub-diffusion regime
has been extremely challenging. Therefore, the experimental
investigation of circularly polarized light in the low-order
scattering (or short traveling photons) regime using enhanced
backscattering (EBS, also known as coherent backscattering) of
light under low spatial coherence illumination will provide a
better understanding of its mechanisms and the polarization
properties of EBS as well.

EBS is a self-interference effect in elastic light scattering,
which gives rise to an enhanced scattered intensity in the
backward direction.  In our previous publications, [5-8] we
demonstrated that low spatial coherence illumination (the spatial
coherence length of illumination $L_{sc}\!<<l_s^*$) dephases the
time-reversed partial waves outside its finite coherence area,
rejecting long traveling waves in weakly scattering media.  EBS
under low spatial coherence illumination ($L_{sc}\!<<l_s^*$) is
henceforth referred to as low-coherence EBS (LEBS).  The angular
profile of LEBS, $I_{LEBS}(\theta)$, can be expressed as an
integral transform of the radial probability distribution $P(r)$
of the conjugated time-reversed light paths:[6-8]
\begin{equation}
I_{LEBS}(\theta)\propto \int^\infty_0 C(r)rP(r)\exp(i2\pi r \theta
/ \lambda)dr,
\end{equation}
where $r$ is the radial distance from the first to the last points
on a time-reversed light path and $C(r)
=|2J_1(r/L_{sc})/(r/L_{sc})|$ is the degree of spatial coherence
of illumination with the first order Bessel function $J_1$.[9] As
$C(r)$ is a decay function of $r$, it acts as a spatial filter,
allowing only photons emerging within its coherence areas ($\sim
L_{sc}^2$ ) to contribute to $P(r)$.  Therefore, LEBS provides the
information about $P(r)$ for a small $r$ ($<\sim100~\mu m$) that
is on the order of $L_{sc}$ as a tool for the investigation of
light propagation in the sub-diffusion regime.

\begin{figure}
\includegraphics[scale=0.47]{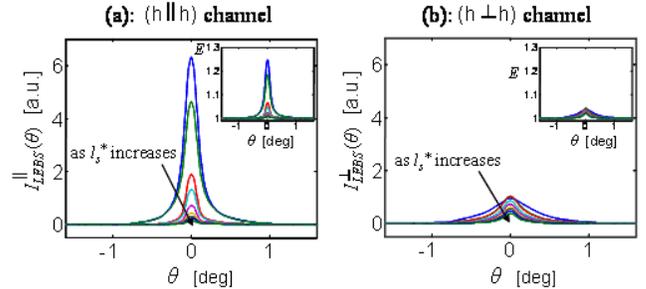}
\caption{Representative $I_{LEBS}(\theta)$ with $L_{sc} = 110~\mu
µm$ obtained from the suspensions of microspheres ($a = 0.15~\mu
m$, $ka = 2.4$, and $g = 0.73$).  We obtained $I_{LEBS}(\theta)$
for various $l_s^* = 67 - 1056 ~\mu m$ ($l_s = 18 - 285 ~\mu m$)
from the (h$||$h) and (h$\bot$h) channels. The insets show the
enhancement factors $E$. }
\end{figure}

To investigate the helicity preservation of circularly polarized
light in the sub-diffusion regime by exploiting LEBS, we used the
experimental setup described in detail elsewhere.[5,6]  In brief,
a beam of broadband cw light from a 100 W xenon lamp
(Spectra-Physics Oriel) was collimated using a 4-$f$ lens system,
polarized, and delivered onto a sample with the illumination
diameter of  $3~mm$.  By changing the size of the aperture in the
4-$f$ lens system, we varied spatial coherence length $L_{sc}$ of
the incident light from $35~\mu m$ to $200~\mu m$.  The temporal
coherence length of illumination was $0.7~\mu m$ with the central
wavelength = $520~nm$ and its FWHM = $135~nm$.  The circular
polarization of LEBS signals was analyzed by means of an
achromatic quarter-wavelet plate (Karl Lambrecht) positioned
between the beam splitter and the sample.  The light backscattered
by the sample was collected by a sequence of a lens, a linear
analyzer (Lambda Research Optics), and a CCD camera (Princeton
Instruments).  We collected LEBS signals from two different
circular polarization channels: the helicity preserving (h$||$h)
channel and the orthogonal helicity (h$\bot$h) channel.  In the
(h$||$h) channel, the helicity of the detected circular
polarization was the same as that of the incident circular
polarization.  In the (h$\bot$h) channel, the helicity of the
detected circular polarization was orthogonal to that of the
incident circular polarization.

In our experiments, we used media consisting of aqueous
suspensions of polystyrene microspheres ($n_{sphere} = 1.599$ and
$n_{water} = 1.335$ at $520~nm$) (Duke Scientific) of various
radii $a$ = 0.05, 0.10, 0.15, 0.25, and 0.45 $\mu m$ (the size
parameter $ka = 0.8 - 7.2$ and the anisotropic factor $g = 0.11-
0.92$).  The dimension of the samples was $\pi \times 252~mm^2
\times 50~mm$. Using Mie theory,[10] we calculated the optical
properties of the samples such as the scattering mean free
pathlength of light in the medium $l_{s}$ ($= 1/\mu_s$, where
$\mu_s$ is the scattering coefficient), the anisotropy factor $g$
(= the average cosine of the phase function), and the transport
mean free pathlength $l_{s}^*$ ($= 1/\mu_s^* = l_{s}/(1 - g)$,
where $\mu_s^*$ is the reduced scattering coefficient).  We also
varied $L_{sc}$ from 40 to 110 $\mu m$. We used $g$ as a metric of
the tendency of light to be scattered in the forward direction.

\begin{figure}
\includegraphics[scale=0.43]{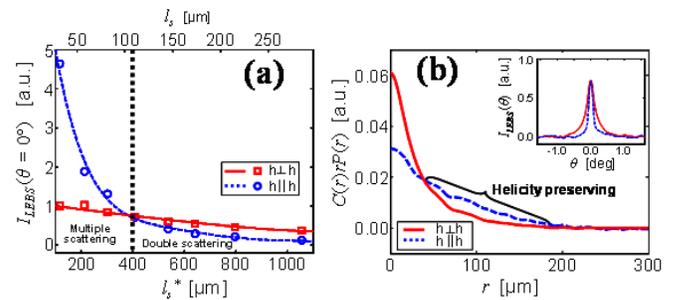}
\caption{$I_{LEBS}$ in the backward direction from Fig. 1. (a)
$I_{LEBS}^{||}(\theta = 0)$ and $I_{LEBS}^{\bot}(\theta = 0)$
cross over at $l_s^* = 408 ~\mu m$ ($l_s = 110~\mu m m$).  The
lines are third-degree polynomial fitting. (b) Inset:
$I_{LEBS}^{||}(\theta)$ and $I_{LEBS}^{\bot}(\theta)$   at the
cross point. $C(r)rP(r)$ obtained by calculating the inverse
Fourier transform of $I_{LEBS}(\theta)$  reveals helicity
preserving in the (h$||$h) channel when $r > \sim50~\mu m$. }
\end{figure}

The total experimental backscattered intensity $I_{T}$   can be
expressed as $I_T = I_{SS} + I_{MS} + I_{EBS}$, where $I_{SS}$,
$I_{MS}$, and $I_{EBS}$ are the contributions from single
scattering, multiple scattering, and interference from the
time-reserved waves (i.e., EBS), respectively.  In media of
relatively small particles (radius, $a\leq\lambda$), the angular
dependence of $I_T(\theta)$ around the backward direction is
primarily due to the interference term, while the multiple and
single scattering terms have weaker angular dependence.
Thus,
$I_{SS} + I_{MS}$ ($=$ the baseline intensity) can be
measured at large backscattering angles ($\theta > 3^{\circ}$).
Conventionally, the enhancement factor $E = 1 +
I_{EBS}(\theta=0^{\circ})/(I_{SS}+I_{MS})$ is commonly used.
However, in the studies of circularly polarized light, the
enhancement factor should be modified, because the intensity of
multiple scattering can be different in the two different channels
and because in the (h$||$h) channel, single scattering is
suppressed due to the helicity flip. Thus, in our studies, we
calculated $I_{EBS}$ by subtracting $I_{SS} + I_{MS}$ from $I_T$.

Figure 1 shows representative LEBS intensity profiles
$I_{LEBS}(\theta)$ from the suspension of the microspheres with
$a$ = 0.15 $\mu m$ ($ka = 2.4$ and $g = 0.73$ at $\lambda =
520~nm$). $I_{LEBS}^{||}$ and $I_{LEBS}^{\bot}$ denote from the
(h$||$h) and (h$\bot$h) channels, respectively. We varied
$l_{s}^*$ from 67 to 1056 $\mu m$ ($l_s$ from 18 to 285 $\mu m$)
with $L_{sc} = 110~ \mu m$. In Fig. 2(a), we plot as a function of
$l_s^*$ (the lines are third-degree polynomial fitting), showing
two characteristic regimes: (i) the multiply scattering regime
($L_{sc} \gg l_s^*$) and (ii) the minimally scattering regime
($L_{sc} \ll l_s^*$).  As expected, in the multiply scattering
regime (i), $I_{LEBS}^{||}$ is higher than $I_{LEBS}^{\bot}$
because of the reciprocity principle in the (h$||$h) channel.  On
the other hand, in the minimal scattering regime (ii), a priori
surprisingly, $I_{LEBS}^{||}$  is lower than $I_{LEBS}^{\bot}$ .
This is because in this regime, LEBS originates mainly from the
time-reversed paths of the minimal number of scattering events in
EBS (i.e., mainly double scattering) in a narrow elongated
coherence volume.[8] In this case, the direction of light
scattered by one of the scatterers should be close to the forward
direction, while the direction of the light scattered by the other
scatterer should be close to the backscattering that flips the
helicity of circular polarization. After the cross point, the
difference between   and remains nearly constant, indicating that
LEBS reaches to the asymptotic regime of double scattering.

More importantly, Fig. 2(a) shows that  and  $I_{LEBS}^{||}$ and
$I_{LEBS}^{\bot}$ cross over at $l_s^* = 408 \mu m$ ($l_s = 110
\mu m$). The cross point can be understood in the context of the
circular polarization memory effect as follows.  As shown in the
inset of Fig. 2(b), at the cross
point,$\int^\infty_0C(r)P^{||}(r)=\int^\infty_0C(r)P^{\bot}(r)$,
where $P^{||}(r)$ and $P^{\bot}(r)$ are the radial intensity
distributions of the (h${||}$h) and (h${\bot}$h) channels,
respectively. Thus, the cross point $R_i$ determines the optical
properties ($l_s^*$ or $l_s$) such that $\int^{\sim
L_{sc}}_0P^{||}(r)=\int^{\sim L_{sc}}_0P^{\bot}(r)$. In other
words, $R_i$ defines $l_s^*$ or $l_s$ such that and are equal
within $L_{sc}$ and thus, the degree of circular polarization
within $L_{sc}$ becomes zero as well. As shown in Fig. 2(b), the
$C(r)rP(r)$, which can be obtained by the inverse Fourier
transform of $I_{LEBS}(\theta)$  using Eq. (1), reveals more
detailed information about the helicity preservation. For small
$r$, $P^{||}(r) < P^{\bot}(r)$. For $r > \sim50~\mu m$, ($\sim
l_s/2$), $P^{||}(r) > P^{\bot}(r)$, showing that the initial
helicity is preserved. This is because the successive scattering
events of the highly forward scatterers direct photons away from
the incident point of illumination, while maintaining the initial
helicity.

\begin{figure}
\includegraphics[scale=0.43]{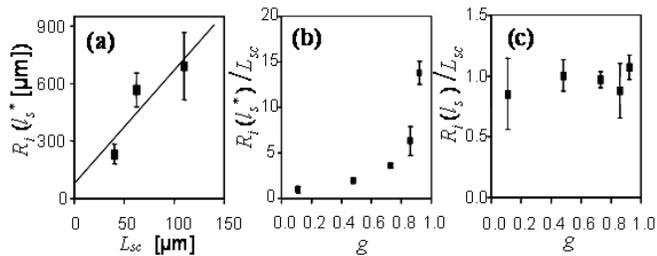}
\caption{Dependence of $R_i$ on $L_{sc}$ and $g$ in LEBS
measurements. (a) Plot of $R_i$ (in the units of $l_s^*$ ) versus
$L_{sc}$ for a fixed $g$ = 0.86 ($ka$ = 4.0).  (b) $R_i$ (in the
units of $l_s^*$)/$L_{sc}$ as a function of $g$. (c) $R_i$ is
recalculated in the units of $l_s$. }
\end{figure}

As discussed above, the cross point $R_i$ is determined by both
the spatial coherence length of illumination $L_{sc}$ and the
optical properties of the media.  Thus, we investigated the
relationship between $L_{sc}$ and $R_i$ using the fixed scatterer
size with $a = 0.25 \mu m$ ($ka = 4.0$, and $g = 0.86$).  Fig.
3(a) shows that $R_i$ (in the units of $l_s^*$) is linearly
proportional to $L_{sc}$ and that small reduced scattering
coefficients $\mu_s^*$ ($= 1/l_s^*$) are necessary to reach a
cross point as $L_{sc}$ increases. Because the linear fitting line
passes through the origin (the 95\% confidence interval of the
intercept of the $L_{sc}$ axis is [$-32~\mu m$, $44~\mu m$]),
$R_i$ can be normalized by $L_{sc}$.  Next, in order to elucidate
how the tendency of the propagation direction (i.e., $g$) plays a
role in the memory effect, we further studied the effect of $g$ on
$R_i$ using the various size parameters $ka$ ranging from 0.8 to
7.2 ($g = 0.11 - 0.92$) with the fixed $L_{sc} = 110~\mu m$.  In
Fig. 3(b), we plot $R_i$  (in the units of $l_s^*$ ) versus $g$.
This shows $R_i$ increases dramatically as $g$ increases, which is
in good agreement with the conventional notion that a small
$\mu_s*$ is required for the memory effect to occur in media of
larger particles because of the stronger memory effect in media of
larger scatterers.  When we plot $R_i$ in the units of $l_s$
versus $g$, as shown in Fig. 3(c), on the other hand, $R_i$ does
not depend strongly on $g$. This result shows that when $l_s$ is
on the order of $L_{sc}$, the helicity of circular polarization is
maintained over a large range of the size parameters.  Moreover,
Fig. 3(c) demonstrates that the average distance of single
scattering events (i.e., $l_s$) is a main characteristic length
scale that plays major roles in the memory effect in the
sub-diffusion regime.

In summary, we experimentally investigated for the first time the
circular polarization memory effect in the sub-diffusion regime by
taking advantage of LEBS, which suppresses time-reserved waves
beyond the spatial coherence area; and thus isolates low-order
scattering in weakly scattering media.  We reported that LEBS
introduces the new length scale (i.e., cross point) at which the
degree of circular polarization becomes zero; and the scale is
determined by both the spatial coherence length of illumination
and the optical properties of the media.  Using the cross point of
the LEBS measurements from the (h$||$h) and (h$\bot$h) channels,
we further elucidated the memory effect in the sub-diffusion
regime. Our results demonstrate that the memory effect exists in
the EBS phenomenon.  Furthermore, we show that the cross point is
the transition point from multiple scattering to double scattering
events this regime.  Finally, our results will further facilitate
the understanding of the propagation of circularly polarized light
in weakly scattering media such as biological tissue.



\bibliography{Cir_pol_memo}
-----------------------------------------------\\
1. F. C. Mackintosh, J. X. Zhu, D. J. Pine, and D. A. Weitz,
``Polarization memory of multiply scattered light,'' Phys. Rev. B 40, 9342 (1989).\\
2. D. Bicout, C. Brosseau, A. S. Martinez, and J. M. Schmitt,
``Depolarization of Multiply Scattered Waves by Spherical Diffusers
- Influence of the Size Parameter,'' Phys. Rev. E 49, 1767 (1994).\\
3.  A. D. Kim and M. Moscoso, ``Backscattering of circularly
polarized pulses,'' Opt. Lett. 27, 1589 (2002).\\
4. M. Xu and R. R. Alfano, ``Circular polarization memory of
light,'' Phys. Rev. E 72, 065601(R) (2005).\\
5.  Y. L. Kim, Y. Liu, V. M. Turzhitsky, H. K. Roy, R. K. Wali,
and V. Backman, ``Coherent Backscattering Spectroscopy,'' Opt. Lett. 29, 1906 (2004).\\
6. Y. L. Kim, Y. Liu, R. K. Wali, H. K. Roy, and V. Backman,
``Low-coherent backscattering spectroscopy for tissue
characterization,'' Appl. Opt. 44, 366 (2005).\\
7. Y. L. Kim, Y. Liu, V. M. Turzhitsky, R. K. Wali, H. K. Roy, and
V. Backman, ``Depth-resolved low-coherence enhanced
backscattering,'' Opt. Lett. 30, 741 (2005). \\
8. Y. L. Kim, P. Pradhan, H. Subramanian, Y. Liu, M. H. Kim, and
V. Backman, ``Origin of low-coherence enhanced backscattering,''
Opt. Lett. 31, 1459 (2006). \\
9.  M. Born and E. Wolf, Principles of optics: electromagnetic
theory of propagation, interference and diffraction of light, 7th
ed. (Cambridge University Press, Cambridge; New York, 1999).\\
10. H. C. van de Hulst, Light scattering by small particles
(Dover Publications, New York, 1995). \\

\end{document}